\documentclass[letterpaper,english,aps,prb,amsmath,amssymb,reprint,author-year,author-numerical,floatfix,raggedfooter,superscriptaddress]{revtex4-1}

\usepackage[latin9]{inputenc}
\usepackage{mathtools}
\usepackage{bm}
\usepackage{amsmath}
\usepackage{amssymb}
\usepackage{graphicx}

\usepackage[caption=false]{subfig}

\usepackage{verbatim}
\makeatletter
\newcommand*{\rom}[1]{\expandafter\@slowromancap\romannumeral #1@}
\makeatother

\makeatletter

\usepackage{dcolumn}
\usepackage{bm}
\usepackage{booktabs}

\usepackage{xcolor}
\usepackage{soul}

\makeatother

\usepackage{babel}
\begin{document}

\title{Monte Carlo analysis of phosphorene nanotransistors}

\author{Gautam~Gaddemane}

\affiliation{Department of Materials Science and Engineering, The University of
Texas at Dallas~\\
 800 W. Campbell Rd., Richardson, TX 75080, USA}

\author{Maarten~L.~{Van~de~Put}}

\affiliation{Department of Materials Science and Engineering, The University of
Texas at Dallas~\\
 800 W. Campbell Rd., Richardson, TX 75080, USA}

\author{William~G.~Vandenberghe}

\affiliation{Department of Materials Science and Engineering, The University of
Texas at Dallas~\\
 800 W. Campbell Rd., Richardson, TX 75080, USA}

\author{Edward~Chen}
\affiliation{Corporate Research, Taiwan Semiconductor Manufacturing Company Ltd. 168, Park Ave. II,\\
Hsinchu Science Park, Hsinchu 300-75, Taiwan, Republic of China}
 
\author{Massimo~V.~Fischetti}

\affiliation{Department of Materials Science and Engineering, The University of
Texas at Dallas~\\
 800 W. Campbell Rd., Richardson, TX 75080, USA}
\email{max.fischetti@utdallas.edu.}

\date{\today}

\begin{abstract}
 Experimental studies on two-dimensional (2D) materials are still in the early stages, and most of the theoretical studies performed to screen these materials are limited to the room-temperature carrier-mobility in the free standing 2D layers.
With the dimensions of devices moving towards nanometer-scale lengths, the room-temperature carrier-mobility --- an equilibrium concept --- may not be the main quantity that controls the performance of devices based on these 2D materials, since electronic transport occurs under strong off--equilibrium conditions. Here we account for these non-equilibrium conditions and, for the case of monolayer phosphorene (monolayer black phosphorus),  show the results of device simulations for a short channel n-MOSFET, using the Monte Carlo method coupled with the Poisson equation, including full bands and full electron-phonon matrix elements obtained from density functional theory.   
Our simulations reveal significant intrinsic limitations to the performance of phosphorene as a channel material in nanotransistors.
\end{abstract}

\maketitle

\section{Introduction}

	Over the last decade and a half, two-dimensional (2D) materials have been widely studied for their potential application in nano-scaled field-effect transistors (FETs), due to their ability to confine carriers to atomically thin layers  providing excellent gate-control\cite{Xu_2017a, Kang_2014} and reduced short-channel effects\cite{Schwierz_2015}. In addition, deviations from ideality,  such as surface roughness, dangling bonds, and interface states, which affect the performance of devices, can be reduced or eliminated thanks to their layered nature. Graphene\cite{Geim_2007, Bolotin_2008,Geim_2008}, silicene\cite{Vogt_2012,Roome_2014,Tao_2015,Li_2013} , silicane\cite{Houssa_2011,Restrepo_2014,Low_2014,Khatami_2019}, germanene\cite{Houssa_2011,Roome_2014,Davila_2014}, phosphorene\cite{Carvalho_2016,Cao_2015,Gaddemane_2018b,Liu_2014exp,Liu_2015}, and monolayer transition metal dichalcogenides (TMD)\cite{Mak_2010,Smets_2019,Larentis_2012,Gopalan_2019,Jin_2014} are some of the most extensively studied 2D materials. \\

Monolayer black phosphorous (phosphorene) has gained significant attention thanks to the high carrier-mobility observed in its bulk counterpart (300-1100~cm$^{2}$/V$\cdot$s for electrons and 150-1300~cm$^{2}$/V$\cdot$s for holes)\cite{Akahama_1983, Morita_1986}. However, experimental information is available almost exclusively for relatively thick many-layer films and is consistent with an anisotropic mobility that decreases quickly with decreasing film thickness\cite{Xia_2014, Li_2014,Liu_2015,Liu_2014exp}. Li $\textit{et al.}$\cite{Li_2014} have observed a thickness dependent hole mobility of around 1000~cm$^{2}$/V$\cdot$s for thick films ($>$ 10 nm), decreasing sharply to 1-10~cm$^{2}$/V$\cdot$s for layers 2-3 nm thin. Dognov $\textit{et al.}$\cite{Doganov_2015} have measured an electron mobility of 189~cm$^{2}$/V$\cdot$s in 10 nm thick films.  For monolayers, the only available result is by Cao $\textit{et al.}$\cite{Cao_2015}, who have reported a mobility of 0.5 and 1~cm$^{2}$/V$\cdot$s for electrons and holes, respectively. 

The literature reports several theoretical calculations of the intrinsic carrier mobility in free-standing phosphorene.  The earlier reported mobility calculations in phosphorene were performed using constant deformation potentials (Bardeen-Shockley deformation potentials\cite{Bardeen_1950}) for scattering rates, and the Takagi formula\cite{Takagi_1994} for the mobility calculations, and predicted a very high electron mobility, ranging from 700 to 26000~cm$^{2}$/V$\cdot$s\cite{Rudenko_2016,Trushkov_2017,Qiao_2014}. The Takagi formula considers only acoustic phonons (single branch only), constant deformation potentials calculated from strain, and ignores intervalley and optical phonon scattering. This oversimplified model has been shown to grossly overestimate the carrier mobility\cite{Nakamura_2017, Gaddemane_2018b}. In our previous work in Ref.~\onlinecite{Gaddemane_2018b}, we found that the deformation potentials for elastic scattering with acoustic phonons are not constant, and vary with the scattering angle (see Fig.~1 of Ref.~\onlinecite{Gaddemane_2018b}). Using $\it{ab~initio}$ methods, we calculated the carrier-mobility in phosphorene using full electron-phonon matrix elements including scattering with all the phonon modes and scattering processes, and obtained a room-temperature electron mobility of 21~cm$^{2}$/V$\cdot$s and 10~cm$^{2}$/V$\cdot$s in the armchair and zigzag directions, respectively.  

Nevertheless, with the current dimensions of devices moving towards nanometer-scale lengths, the room-temperature carrier-mobility might not be a suitable quantity to determine the full potential of phosphorene as a channel material. In the strong off-equilibrium conditions  that are present in such small devices even at a moderate source-drain bias, charge carriers are driven to energies at which concepts like mobility and effective mass, which are well-defined only at equilibrium and near a band-edge (conduction-band minimum or valence-band maximum), lose their meaning and predictive power\cite{Fischetti_1991}.
Additional off-equilibrium effects, such as hot-electron effects and velocity overshoot, also become important.  Therefore, to fully determine the potential usefulness of 2D materials in transistor technology, we must employ full device simulations including high energy bands, scattering rates evaluated at higher energies, and models which take into account the non-equilibrium effects.

 Previously-published simulation results based on ballistic quantum transport predict an on-state current of approximately 5500~$\mu$A/$\mu$m  in a 20 nm channel length device\cite{Cao_2014}. When including electron-phonon scattering in the quantum transport simulations, Szabo $\textit{et al.}$\cite{Szabo_2015} predict an on-state current approximately 2624~$\mu$A/$\mu$m for a 10 nm channel device, and Brahma $\textit{et al.}$\cite{Brahma_2019} predict an on-state current approximately 2340~$\mu$A/$\mu$m for a 20 nm channel device. All the above mentioned results were obtained by considering armchair orientation as the transport direction. However and unfortunately, these results are affected by the same crude approximations that we have noticed before regarding calculations of the low-field mobility. Indeed, in Refs.~\onlinecite{Szabo_2015} and~\onlinecite{Brahma_2019}, the authors use constant deformation potentials, evaluated from band-structure calculations of the strained material (Bardeen-Shockley deformation potentials), to treat acoustic scattering. As shown later in the results section, by accounting for the anisotropy of the deformation potentials, scattering with all the phonon branches, and intra- and intervalley scattering processes, the on-current we have calculated  is significantly lower than what has been previously predicted in Refs.~\onlinecite{Cao_2014,Szabo_2015,Brahma_2019}.

In this paper, we present  results of room-temperature device simulations for a double-gate phosphorene nanometer-scale field-effect transistor (FET), which are performed by coupling the Boltzmann transport equation (BTE) with the Poisson equation. We solve the BTE using the Monte Carlo method. We use full bands, and full electron-phonon matrix elements for the first two conduction bands obtained from first-principles methods (density-functional theory, DFT). We include inter- and intra-valley scattering processes assisted by all acoustic and optical phonons. We also take into account degeneracy effects in the highly doped regions.

Since studies of nanometer-scale devices have been published using quantum-transport models in the ballistic limit or with scattering using the deformation potential approximation, here we wish to account for dissipative scattering process in a more rigorous manner. Indeed, it may be argued that scattering controls electronic transport at room temperature even in devices with active regions as short as 5 or 10 nm (experimental evidence of coherent transport at 300 K has never been provided). In this way, we wish to provide an ideal best-case estimate of the performance of phosphorene-based FETs compared to ballistic studies and simulations including scattering under the deformation potential approximation, that also assume ideal free-standing monolayers. Therefore, in the calculations presented here we have ignored additional scattering processes, such as  impurity scattering\cite{ong2014anisotropic}, remote-phonon scattering\cite{fischetti2001effective}, and electron-electron scattering.

This paper is organized as follows: In Section~\ref{s:theory}, we present the general theoretical framework used in our device simulations for 2D materials based FETs. In Section~\ref{s:results}, we show the results of our device simulations on phosphorene FETs. Finally, we present our conclusions in Section~\ref{s:conclude}. \\ 

\section{Theoretical Model}
\label{s:theory}

\begin{figure}[!t]
\centering
\includegraphics[scale=0.9]{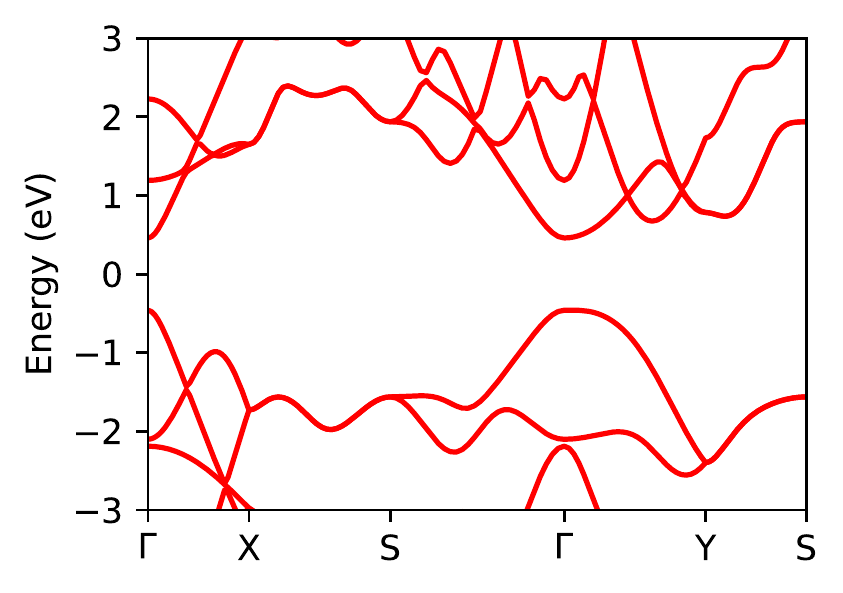}
\caption{The band structure of phosphorene calculated using QE.}
\label{fig:bands}
\end{figure}

\begin{figure}[!t]
\centering
\includegraphics[scale=0.9]{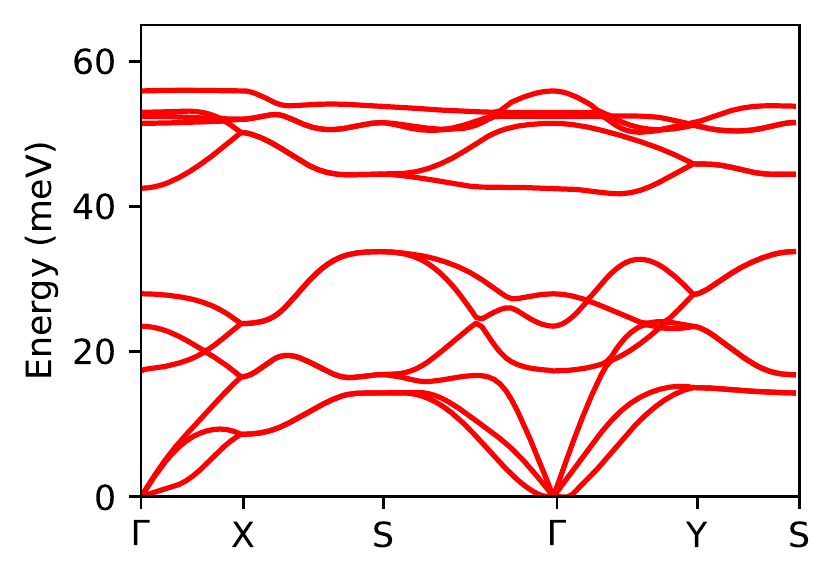}
\caption{The phonon dispersion of phosphorene calculated using QE.}
\label{fig:phonons}
\end{figure}

\begin{figure}[!t]
\centering
\includegraphics[scale=0.9]{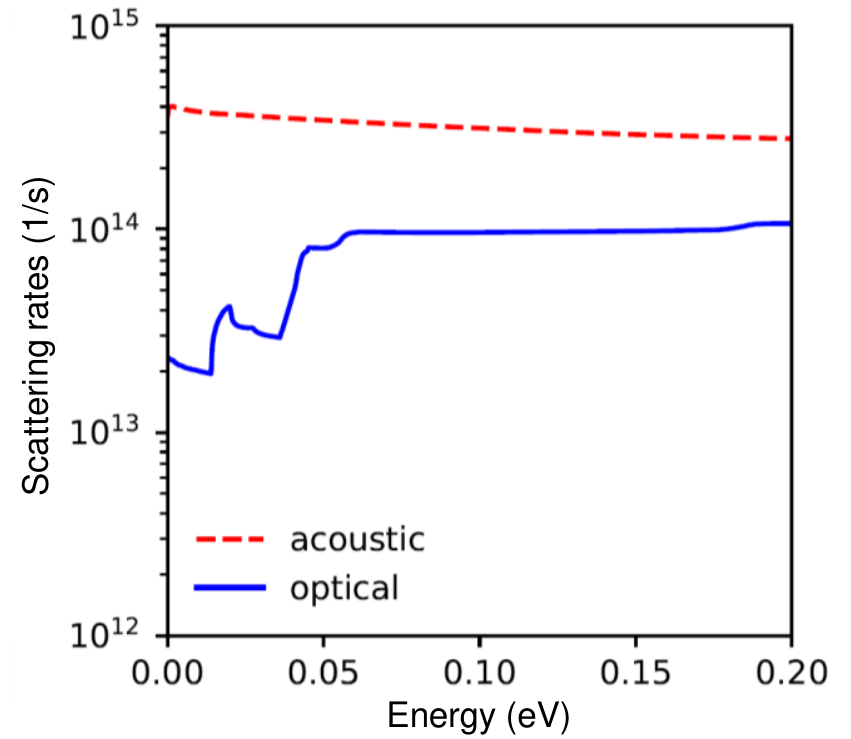}
\caption{The angle-averaged electron-phonon scattering rates as a function of kinetic energy in phoshphorene at 300 K.}
\label{fig:srates}
\end{figure}

In this section, we provide a brief description of the key components of the theoretical model used in our device simulations. The current work in an extension of our previous work on homogeneous transport in phosphorene\cite{Gaddemane_2018b}.  

\subsection{Electronic band structure}

The electronic band structure is obtained from density functional theory (DFT) using the Quantum ESPRESSO  (QE)\cite{Giannozzi_2009} package with the Perdew-Burke-Enzerhoff (PBE) generalized gradient approximation for the exchange-correlation functional, and ultrasoft pseudopotentials. We show in Fig.~\ref{fig:bands}, the electronic band structure of phosphorene calculated from QE. From the figure, we can see that phosphorene has an anisotropic band structure with the conduction band minima located at the $\Gamma$ symmetry point, and an effective mass of 0.14m$_{0}$ along the armchair direction ($\Gamma$-$X$) and 1.24m$_{0}$ along the zigzag direction ($\Gamma$-$Y$), where m$_{0}$ is the free-electron mass. There are two additional valleys, the $Q$ and the $Y$ valley, along the zigzag direction with an energy separation of 0.21~eV and 0.27~eV, respectively, from the conduction band minima at the $\Gamma$ symmetry point. These valleys become important in transport simulations as they become significantly populated at high electric fields (see Fig.~9 of Ref.~\onlinecite{Gaddemane_2018b}).

To account for the full electronic dispersion in our transport simulations, the band structure (E($\bf{K}$)) for the first two conduction bands is tabulated over two nested meshes, a coarse and a fine mesh, in the first quadrant of the rectangular Brillouin zone (BZ). It is necessary to adopt an extremely fine mesh around the conduction-band minima to account for the anisotropy and the strong non-parabolicity of the band structure. In addition, we have calculated and tabulated the gradients ($\nabla_{\bf{K}}E(\bf{K})$) on the same meshes of the BZ, using a finite difference scheme, to obtain the electron velocity. We evaluate the electron energy and velocity for any arbitrary $\bf{K}$ point and band $\eta$  by first mapping the $\bf{K}$ point to the first quadrant, and then linearly interpolate from the nearest mesh points. The sign for the velocity is assigned based on the original quadrant of the $\bf{K}$ point in the BZ. The computational parameters and the details of the meshes used in the current study are given in Ref.~\onlinecite{Gaddemane_2018b}. 

\subsection{Phonon dispersion and electron-phonon scattering rates}

The phonon dispersion and electron-phonon matrix elements are calculated using density functional perturbation theory (DFPT) as implemented in QE\cite{Baroni_2001}, augmented by the Electron Phonon Wannier (EPW) software package\cite{Giustino_2007}. The electron-phonon scattering rates are then calculated using Fermi's Golden rule. We show in Fig.~\ref{fig:phonons}, the phonon dispersion of phosphorene plotted along the principal symmetry directions. There are a total of nine phonon branches, three acoustic and six optical. In Fig.~\ref{fig:srates}, we show the angle-averaged scattering rates as a function of carrier kinetic energy. The label `acoustic' represents the sum of the scattering rates of the in-plane acoustic branches (longitudinal acoustic (LA) and transverse acoustic (TA)), and the label `optical' identifies the sum of the scattering rates of all six optical branches. The in-plane acoustic phonons dominate the scattering processes. The rate for scattering with the out-of-plane acoustic phonons (ZA) is negligible to the first order due to the horizontal mirror symmetry of the phosphorene structure\cite{Fischetti_2016}. Intervalley scattering is dominated by longitudinal optical (LO) phonons (32~meV) with a deformation potential of about 1.7$\times$10$^{9}$~eV/cm. Intervalley scattering affects transport at high bias. The phonon dispersion and the scattering rates are calculated and tabulated on the same meshes as the band structure, and interpolated in a similar way.

\subsection{Monte Carlo transport model}

We perform the device simulations by solving self-consistently the Boltzmann transport equation for two-dimensional electron gas (2D-BTE), and the Poisson equation. The solution of the 2D-BTE is obtained numerically using the `full-band ensemble' Monte Carlo method\cite{Jacoboni_1983} to treat non-equilibrium transport accurately. The Poisson equation is solved in 2D ($\it{x-z}$ plane in Fig.~\ref{fig:dblegate}) assuming that the device is infinitely long in the missing third direction ($\it{y}$-direction in  Fig.~\ref{fig:dblegate}).  There is no significant variation of physical quantity along the $y$ direction. Therefore, it is sufficient to solve the Poisson equation on the $\it{x-z}$ plane. 

 The procedure and the numerical methods are adapted from Refs.~\onlinecite{Damocles} and~\onlinecite{Fischetti_1988}. To give some details about the implementation of the Monte Carlo method we have employed, the motion of an ensemble of `particles' is simulated, subjecting them to an electric field obtained by solving the Poisson equation and given scattering processes. We use statistical enhancement techniques\cite{Damocles} to control the number of particles in different regions of the device, to enhance the statistics when fewer number of particles are present. This becomes extremely important when calculating the current in the sub-threshold regime as very few particles are present in the channel. We found that a target value of approximately 5000 particles in the channel is necessary when calculating the current in the sub-threshold regime. An initial electron state, labeled by a wavevector ${\bf K}$ and position ${\bf R}$,  is assigned to each particle stochastically according to the equilibrium Fermi Dirac distribution and doping profile, respectively. The particles evolve following the equations of motion, which are given by:
\begin{equation}
\frac{d{\bf K}}{dt} = \frac{e}{\hbar}\nabla_{\bf{R}}V({\bf R},z_{\rm layer},t),
\label{eq:k_mot}
\end{equation}
 and
\begin{equation}
\frac{d{\bf R}}{dt} = \frac{1}{\hbar}\nabla_{\bf{K}}E_{n}({\bf K}),
\label{eq:r_mot}
\end{equation}
where $z_{\rm layer}$ is the position in the 2D layer, for a duration of time ${\Delta t}$. The choice of the time-step ${\Delta t}$ must satisfy several criteria: A correct discretization of the equations of motion, a correct evaluation of the scattering probability at each time step, and a satisfactory representation of the plasma oscillations, as explained in Ref.~\onlinecite{Fischetti_1988} (${\Delta t}$ = 10$^{-16}$ s). At the end of the free flight, a random scattering process is selected stochastically. Based on the chosen scattering process, a new ${\bf K}$ state is chosen at random with probability weighted by the density of final states and squared matrix elements of the scattering process. Stochastic estimators of quantities like kinetic energy, velocity, charge density, electric potential, and current are recorded at the end of each iteration. This procedure is repeated until steady state is reached.

We use the finite difference method to obtain the numerical solution of the Poisson equation. We discretize the $\it {x-z}$ plane (Fig.~\ref{fig:dblegate}) with a square mesh with sides~0.5~nm long. The contacts are defined as boundary conditions for both the BTE and the Poisson equation\cite{Fischetti_1988}. For the BTE, the contact absorbs any particle that attempts to leave the simulated region and particles are injected (with an equilibrium Fermi Dirac distribution in reciprocal space) conditionally in order to maintain charge neutrality at the contact nodes. In solving the Poisson equation, the contacts are treated as boundaries subject to Dirichet boundary conditions. Note that the gate contact is treated as a pure Poisson boundary, since tunneling though the gate dielectric is ignored and no particles can be absorbed or injected by the gate. \\

Degeneracy effects become important in heavily-doped regions. We have incorporated degeneracy effects in the Monte Carlo simulation by updating the distribution function in the reciprocal space (${\bf K}$-space) self-consistently. Given the large amount of memory requirement and detailed statistics that would be necessary to store the full wavevector- and position-dependent distribution function, we follow the procedure employed in Ref.~\onlinecite{Fischetti_1988} and approximate the distribution function with a heated Fermi-Dirac distribution:
\begin{equation}
f({\bf K},{\bf R},\eta,t) = f_{0}({\bf K},\eta,T_{\rm p}({\bf R},t),E_{\rm F}({\bf R},t)),
\label{eq:zeroh}
\end{equation} 

where

\begin{equation}
f_{0}({\bf K},\eta,T,E_{\rm F}) = \bigg({1+\exp\bigg[\frac{E_{\eta}({\bf K})-E_{\rm F}}{K_{\rm b}T}\bigg]}\bigg)^{-1}
\label{eq:zeroh}
\end{equation} 
is the Fermi-Dirac distribution function at temperature T in band $\eta$ and Fermi level E$_{\rm F}$. The local particle temperature $ T_{\rm p}({\bf R},t)$ and Fermi level $E_{\rm F}({\bf R},t)$ are determined in a self-consistent way during the simulation from the local particle density $n({\bf R},t)$ and average kinetic energy $\bar{E}({\bf R},t)$. 
This approximation can be justified in part by the expected strong `thermalization' induced by the strong electron-electron interactions expected at the high carrier density of interest here. Finally, after each scattering event, the collision is accepted or rejected (by leaving the electron state unchanged) stochastically, depending on whether the final state is empty or occupied; that is, dependent on whether $r \geq f$ or $r<f$, respectively. where $r$ is a random number $0\leq r\leq1$, and $f$ is given by Eq.~(\ref{eq:zeroh}) evaluated at the final state ${\bf K}$, band $\eta$  and position ${\bf R}$.

\section{Results and discussion}
\label{s:results}

\begin{figure}[!t]
\centering
\includegraphics[]{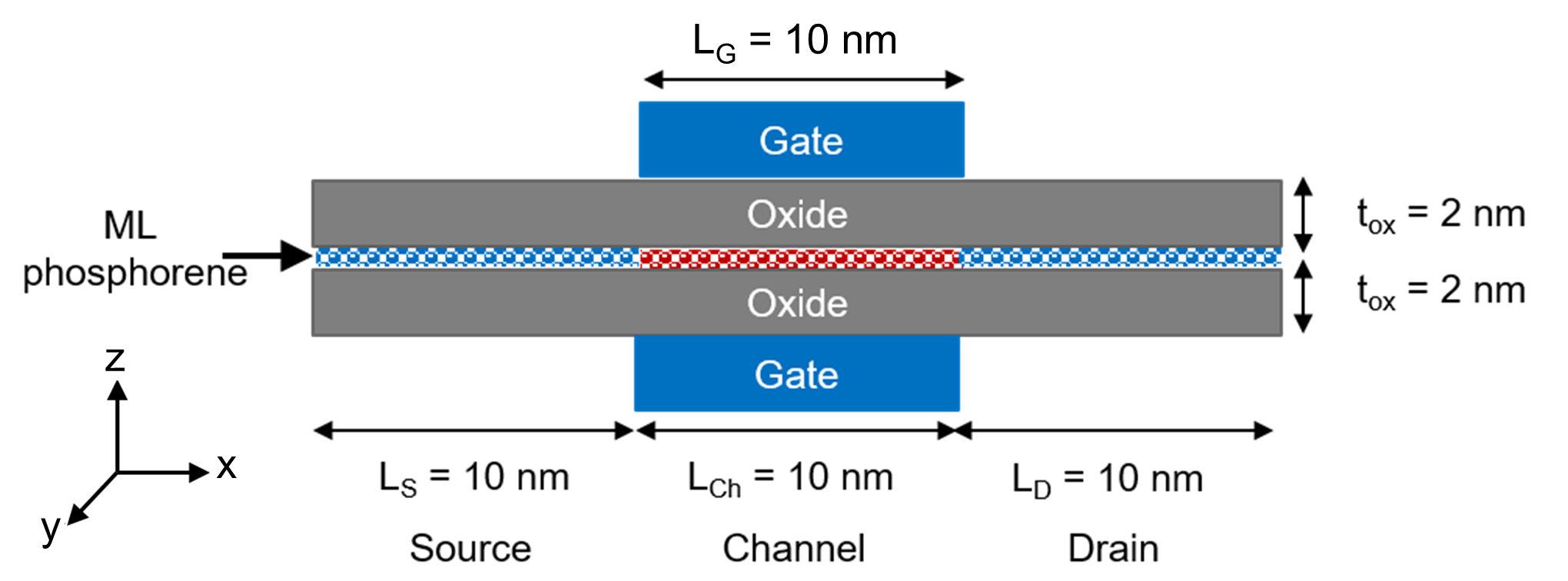}
\caption{Cross section of double-gate phosphorene n-MOSFET used in our simulation.}
\label{fig:dblegate}
\end{figure}

\begin{figure}[!t]
\centering
\includegraphics[scale=0.9]{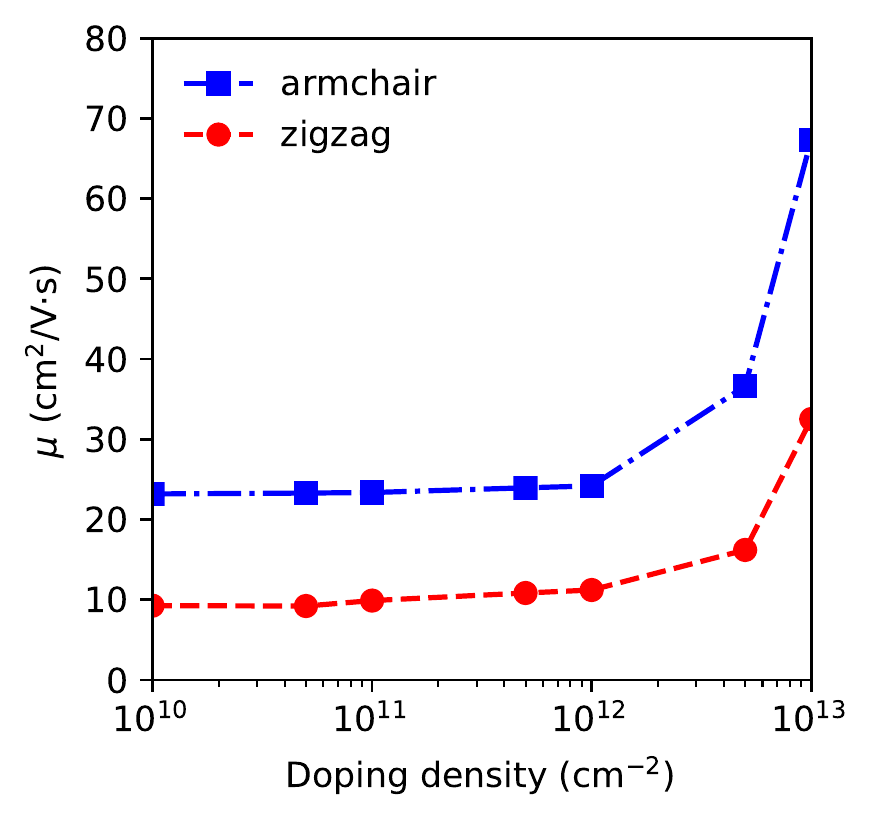}
\caption{Variation of low-field mobility in phosphorene with the doping concentration.}
\label{fig:dop_mob}
\end{figure}

\begin{figure}[!t]
\centering
\includegraphics[scale=0.9]{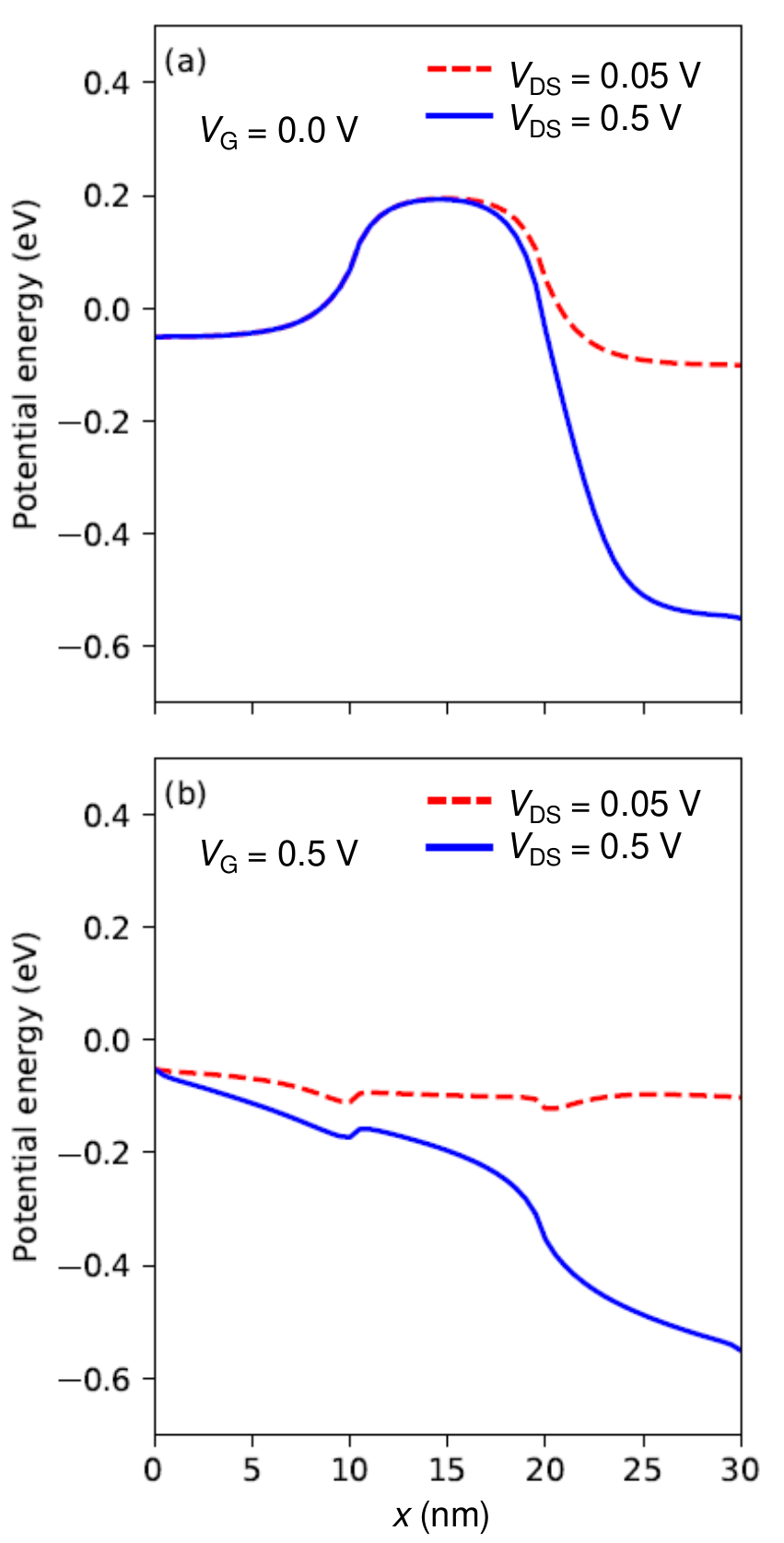}
\caption{Potential energy profile in the 2D layer, plotted along the transport direction for $V_ {\rm DS}$ = 0.05 and 0.5~V at (a) $V_{\rm G}$ = 0.0~V (off-state), and (b)  $V_{\rm G}$ = 0.5~V (on-state).}
\label{fig:pe}
\end{figure}

\begin{figure}[!t]
\centering
\includegraphics[scale=0.9]{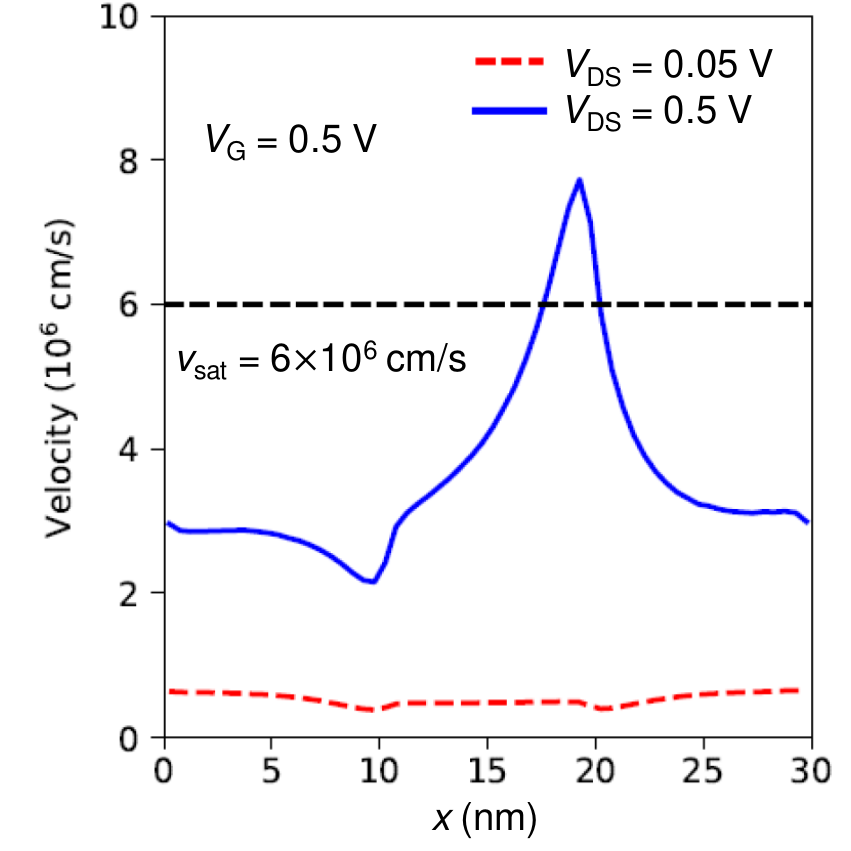}
\caption{Average electron velocity in the 2D layer, plotted along the transport direction for $V_ {\rm DS}$ = 0.05 and 0.5~V at (a) $V_{\rm G}$ = 0.0~V (off-state), and (b)  $V_{\rm G}$ = 0.5~V (on-state).}
\label{fig:bp_ve}
\end{figure}

\begin{figure}[!t]
\centering
\includegraphics[scale=0.9]{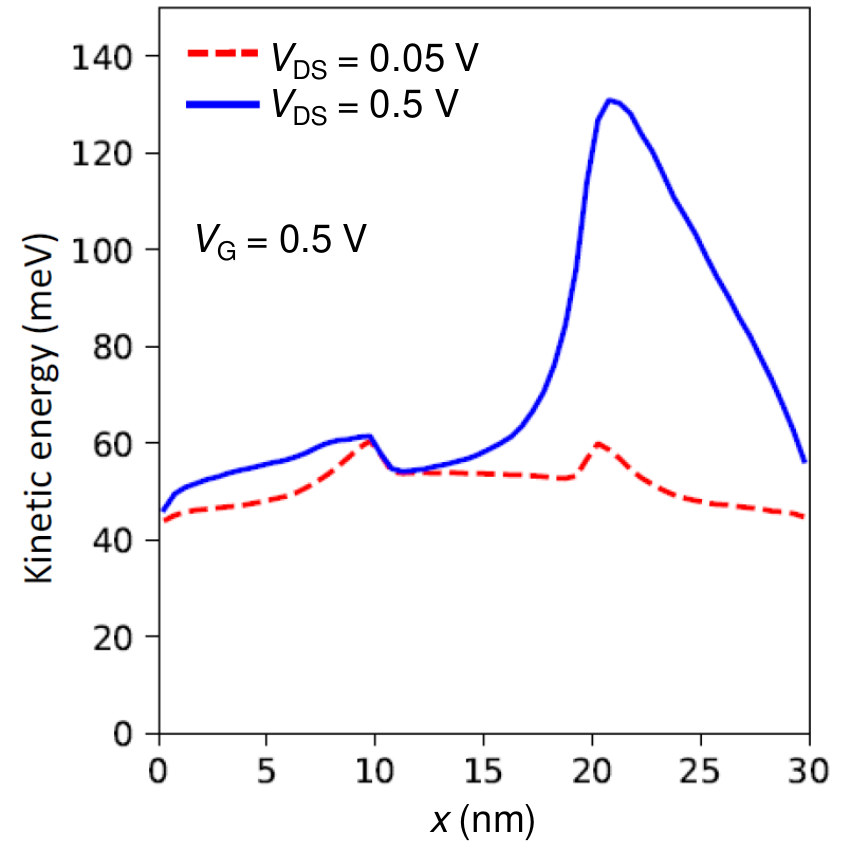}
\caption{Average electron kinetic energy in the 2D layer, plotted along the transport direction for $V_ {\rm DS}$ = 0.05 and 0.5~V at (a) $V_{\rm G}$ = 0.0~V (off-state), and (b)  $V_{\rm G}$ = 0.5~V (on-state).}
\label{fig:bp_en}
\end{figure}

\begin{figure}[!t]
\centering
\includegraphics[]{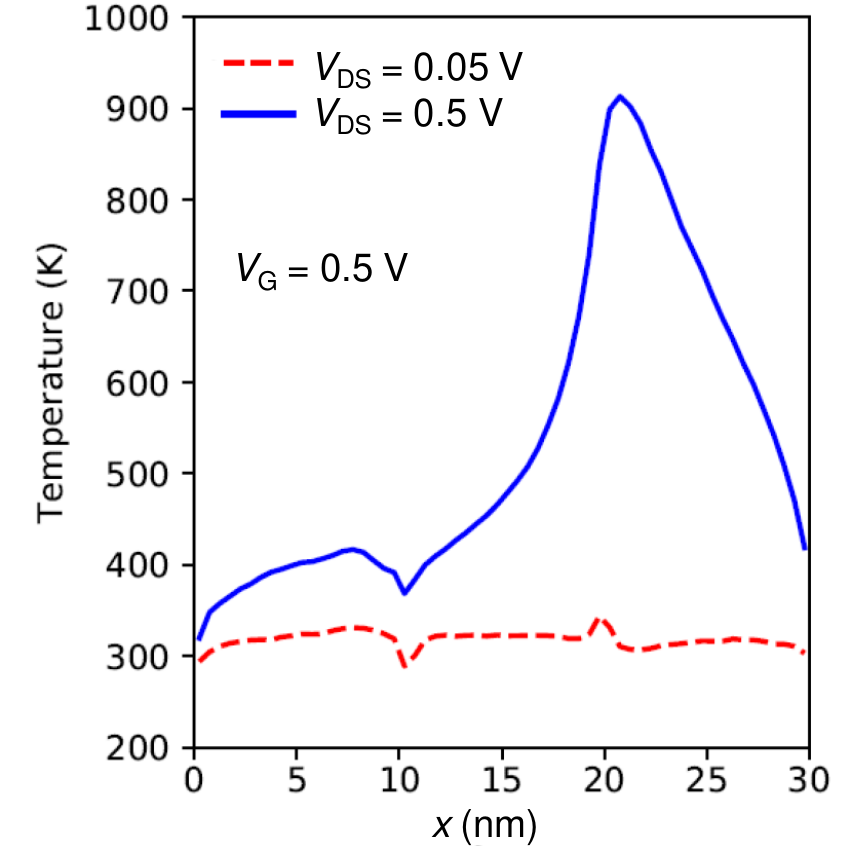}
\caption{Average electron temperature in the 2D layer, plotted along the transport direction for $V_ {\rm DS}$ = 0.05 and 0.5~V at (a) $V_{\rm G}$ = 0.0~V (off-state), and (b)  $V_{\rm G}$ = 0.5~V (on-state).}
\label{fig:bp_temp}
\end{figure}

\begin{figure}[!t]
\centering
\includegraphics[]{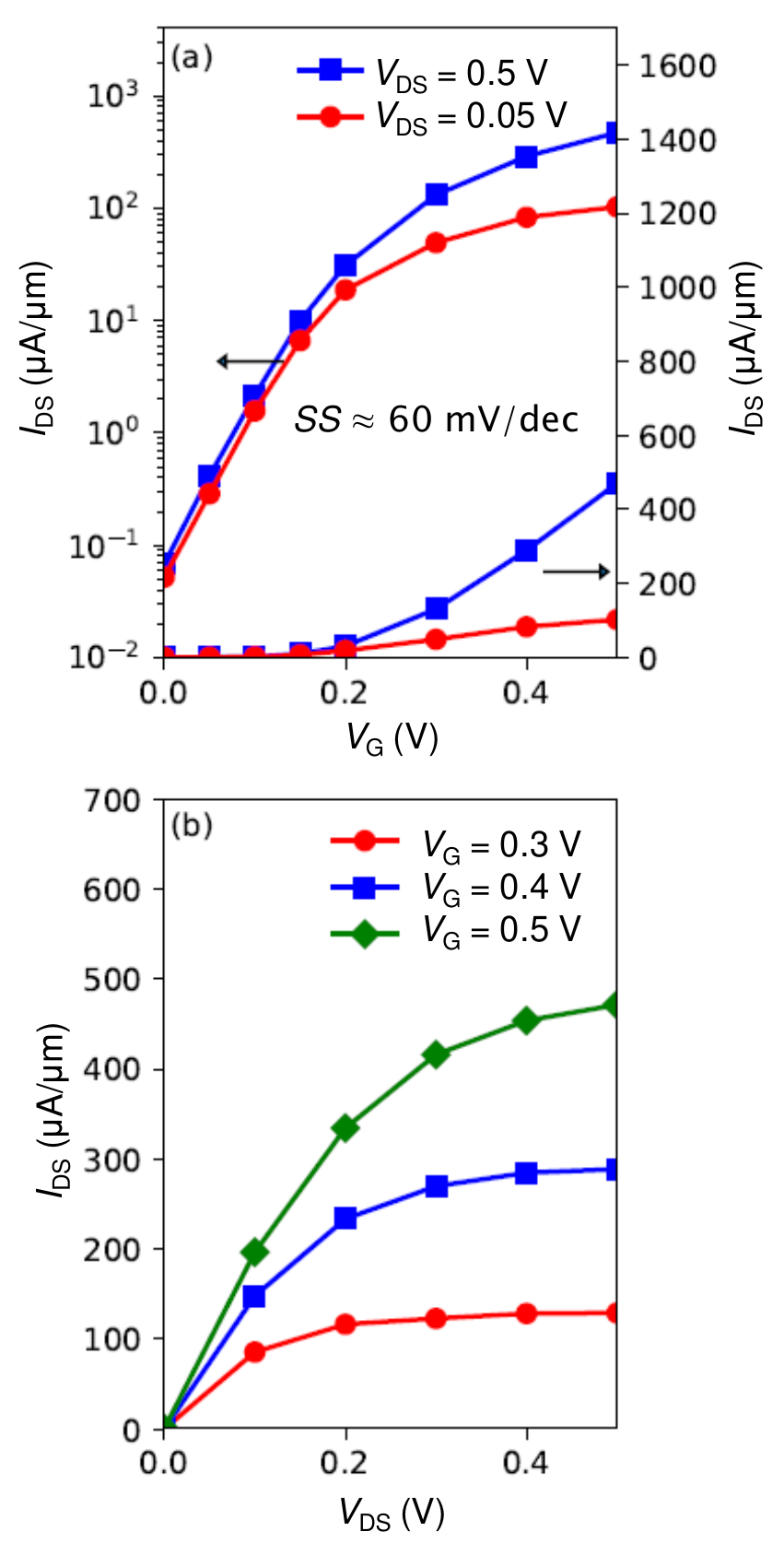}
\caption{(a)Transfer characteristics ($I_{\rm DS}$ ${\it vs.}$ $V_{\rm G}$) plotted on both linear and log scale for $V_{\rm DS}$ = 0.05 and 0.5~V, and (b) output characteristics ($I_{\rm DS}$ ${\it vs.}$ $V_{\rm DS}$) for $V_{\rm G}$ = 0.3, 0.4, and 0.5~V  }
\label{fig:I_vs_V}
\end{figure}

In this section, we start by describing the geometry of the device used in our simulation, followed by a discussion on the physical information extracted from our simulations. Finally we show and discuss the transfer ($I_{\rm DS}$ ${\it vs.}$ $V_{\rm G}$) and output characteristics ($I_{\rm DS}$ ${\it vs.}$ $V_{\rm DS}$).  

In Fig.~\ref{fig:dblegate}, we show the cross section of the double-gate device we simulated. Summarizing the design of the device, the source and drain extensions of length $L_{\rm S/D}$~=~10~nm are doped degenerately with a doping density of 10$^{13}$~cm$^{-2}$. The channel of length $L_{\rm CH}$~=~10~nm is left undoped. We have chosen the armchair direction as the channel orientation, since from previous works~\cite{Gaddemane_2018b,Cao_2015,Szabo_2015,Li_2014}, it has been found that the intrinsic mobility is larger in this direction compared to the zigzag direction. Aluminium oxide (Al$_{2}$O$_{3}$) with a thickness of 2 nm, which corresponds to 0.7 nm effective-oxide thickness (EOT), is used as the gate insulator. For simplicity, we have taken the gate perfectly aligned with the channel without any underlap or overlap. The simulations are performed assuming room-temperature operation.

The source/drain extensions are doped degenerately with a doping density of 10$^{13}$~cm$^{-2}$ to reduce the resistance across the extensions. In these regions the large carrier density yields very low-fields and negligible carrier heating. Therefore, the resistance (voltage drop) depends on the low-field mobility. Figure~\ref{fig:dop_mob} shows the variation of the carrier mobility, along both the armchair and zigzag direction, with the doping concentration. The carrier mobility increases with increasing doping density thanks to degeneracy effects. For degenerately-doped materials, the scattering rates are reduced by the smaller number of available (empty) final states (Pauli blocking). At a doping density of 10$^{13}$~cm$^{-2}$, the low-field mobility in the armchair direction increases from its intrinsic value of 21~cm$^{2}$/V$\cdot$s to 70~cm$^{2}$/V$\cdot$s. However, in practice, other effects --- ignored here --- may result in a qualitatively different outcome. For example, if the doping is obtained via impurities, impurity scattering will matter.  

In Fig.~\ref{fig:pe}, we show the potential-energy profile in the 2D layer, plotted along the transport direction for 
$V_ {\rm DS}$~=~0.05~V and 0.5~V at $V_{\rm G}$ = 0.0~V (off-state), and $V_{\rm G}$ = 0.5~V (on-state). In the off-state 
(Fig.~\ref{fig:pe}a), we observe the absence of any appreciable drain-induced barrier-lowering (DIBL) by increasing $V_ {\rm DS}$ from 0.05~V to 0.5~V, thanks to the double-gate structure. In the on-state, at  $V_ {\rm DS}$ = 0.5~V, we observe a significant potential drop in the source/drain extensions (0--10 nm and 20--30 nm) compared to the potential drop in the channel (10--20 nm). This can be attributed to the very low mobility of phosphorene. Despite the degenerate doping in the extension regions, which also increases the electron mobility, the resistance across the extensions is still quite large. 

 In Fig.~\ref{fig:bp_ve}, we show the average velocity of electrons in the 2D layer, plotted along the transport direction for $V_ {\rm DS}$ = 0.05~V and 0.5~V at $V_{\rm G}$ = 0.5 V. The average velocity remains constant at the source/drain extensions and reaches a peak value of  approximately 8$\times$10$^{6}$~cm/s for $V_ {\rm DS}$~=~0.5~V. In our previous work (see Fig.~12a of Ref.~\onlinecite{Gaddemane_2018b}) we observed a saturation velocity $v_{\rm sat}$ $\approx$ 6$\times$10$^{6}$~cm/s. Therefore, definite velocity overshoot occurs at the drain end of the device.

In Fig.~\ref{fig:bp_en}, we show the average kinetic energy of electrons in the 2D layer, plotted along the transport direction  for $V_ {\rm DS}$ = 0.05~V and 0.5~V at $V_{\rm G}$ = 0.5~V. The average kinetic energy at the source/drain extensions remain close to approximately 50 meV, which is the average equilibrium energy of electrons at room temperature for a doping density of 10$^{13}$ ~cm$^{-2}$. However, for $V_ {\rm DS}$ = 0.5~V, the kinetic energy reaches a peak value of approximately 130 meV at the drain end of the channel. This shows that the electrons become significantly hot at the drain end. From Fig.~\ref{fig:bp_temp}, we can see that the electron temperature reaches a peak value of 900 K at the drain end of the channel.

Finally, In Fig.~\ref{fig:I_vs_V}, we show the transfer characteristics (Fig.~\ref{fig:I_vs_V}a) calculated for $V_{\rm DS}$ = 0.05~V and 0.5 V, and output characteristics (Fig.~\ref{fig:I_vs_V}b) for $V_{\rm G}$ = 0.3, 0.4 and 0.5 V. We obtain an $I_{\rm ON}$ $\approx$ 472 $\mu$A/$\mu$m at $V_{\rm DS}$~=~0.5~V, a sub-threshold slope close to 60~mV/dec, and an on-off ratio of approximately 10$^{3}$. We observe a transconductance $g_{\rm m}$ $\approx$~1622~S/m. 

The on-current we obtain is at least one order of magnitude lower than the theoretical values previously published (Refs.~\onlinecite{Cao_2014,Szabo_2015,Brahma_2019}). As we have discussed in the Introduction section, the authors of Ref.~\onlinecite{Cao_2014} report an $I_{\rm ON}$~$\approx$~5500~$\mu$A/$\mu$m in the ballistic limit for a 20 nm channel length device. As shown above, scattering is extremely important, and the ballistic limit should be considered as a best case scenario, probably excessively optimistic in a low-mobility material such as monolayer phosphorene. The authors of Refs.~\onlinecite{Szabo_2015} and \onlinecite{Brahma_2019} include electron-phonon  scattering in their simulations, and obtain an  $I_{\rm ON}$~$\approx$~2624~$\mu$A/$\mu$m for a 10 nm channel-length device, and $I_{\rm ON}$~$\approx$~2340~$\mu$A/$\mu$m for a 20 nm channel length device, respectively. However, they treat the scattering with acoustic phonons using deformation potentials calculated from strain (Bardeen-Shockley approach). As mentioned above, the deformation potential for acoustic phonons are anisotropic and ignoring them would lead to an overestimation of the transport properties. In our previous work (see Fig.~1 of Ref.~\onlinecite{Gaddemane_2018b}), we found that the deformation potential for acoustic phonons varies from 2~eV to 20 eV with scattering angle, even exhibiting a significant probability for backscattering, an effect that affects electron transport very severely. In addition, the electrons can reach high energies in the channel due to the large fields present and dissipative scattering processes,  like inelastic acoustic scattering, optical phonon scattering, and intervalley scattering become important.

\section{Conclusion}
\label{s:conclude}
 We have simulated a double-gate phosphorene nanotransistor using a full-band ensemble Monte Carlo method. We have used the full band structure and electron-phonon matrix elements obtained from $\it{ab~initio}$ (DFT) calculations in order to account correctly for high-energy effects, for collisions with all phonon branches, and for the anisotropy of the scattering processes. We found that scattering has a significant impact on the performance of the device, and we obtain a much lower on-current than previously reported simulation results. We found that the on-current is significantly affected by the resistance across the source/drain extension. The performance can be enhanced by increasing the doping across the extensions or by reducing the length of the extensions. 

The experimental results available are mostly limited to thick films or many layers of black-phosphorous, and for long channel devices. However, It is evident that the performance of the device decreases with the thickness of the film. 
Although, to obtain results comparable to experimental values, additional scattering processes such as impurity scattering, remote-phonon scattering, Coulomb scattering need to be included, the intrinsic performance of phosphorene FETs  already falls below the ITRS requirement for 2D FETs\cite{ITRS} ($I_{\rm ON}$~$\approx$~2087~$\mu$A/$\mu$m at $V_{\rm DS}$~=~0.62~V and $L_{\rm G}$~=~8.1 nm). In addition, phosphorene is reported to be unstable under ambient conditions\cite{pei2016producing}. Considering these factors  --- together with the observation that the on-current we have calculated satisfies only marginally (if at all) future needs of the VLSI industry --- we suspect that phosphorene might not be a suitable candidate for a channel material in 2D-FETs.

 Only using a more complete and realistic model one can make quantitative predictions about the best possible intrinsic performance of 2D-FETs and whether or not, withstanding processing difficulties, they may be regarded as possible competitors to Si-based technology at the 5--10 nm length. Ballistic transport may be excessively optimistic for low-mobility (high-scattering rates) materials like phosphorene  even for sub-10 nm devices. Therefore, with scattering still playing an important role in short-channel 2D devices, the Monte Carlo method (semiclassical transport) provides an effective avenue to screen and study various 2D materials for CMOS application.

\acknowledgments{This work has been supported in part by the Semiconductor Research Corporation (SRC nCORE) and Taiwan Semiconductor Manufacturing Company, Ltd (TSMC).}

\bibliographystyle{apsrev4-1}
\bibliography{bp_device}

\end{document}